\documentstyle[mnextra]{mn}
\input{epsf}
\seceqnum           
\def\etal{{\rm et al.} }
\def\ie {{\rm i.e. }}
\def\eg {{\rm e.g. }}
\def\log {{\rm log}}
\def\c {{\rm c}}
\def\d {{\rm d}}
\def\t {{\rm t}}

\def\j {{\rm j}}
\def\bfx {{\bf x}}
\def\bfy {{\bf y}}
\def\bfk {{\bf k}}
\def\bfz {{\bf z}}
\def\that {\hat T}
\def\lsim{\mathrel{\hbox{\rlap{\hbox{\lower4pt\hbox{$\sim$}}}\hbox{$<$}}}}
\def\gsim{\mathrel{\hbox{\rlap{\hbox{\lower4pt\hbox{$\sim$}}}\hbox{$>$}}}}

\begin{document}
\title[Speedy pixon image reconstruction]
{A speedy pixon image reconstruction algorithm}
\author[V. Eke]{Vincent Eke$^{1,2}$ \\
$^1${Institute of Astronomy, Madingley Road, Cambridge. CB3 0HA, UK}\\
$^2${Steward Observatory, 933 N Cherry Ave, Tucson. AZ 85721, USA}\\
}

\maketitle

\begin{abstract}
A speedy pixon algorithm for image reconstruction is described.
Two applications of the method to simulated astronomical data sets are
also reported. In one case, galaxy clusters are extracted from
multiwavelength microwave sky maps using the spectral dependence of the
Sunyaev-Zel'dovich effect to distinguish them from the microwave
background fluctuations and the instrumental noise. The second example
involves the recovery of a sharply peaked emission profile, such as
might be produced by a galaxy cluster observed in X-rays.
These simulations show the ability of the technique both to
detect sources in low signal-to-noise data and to deconvolve a
telescope beam in order to recover the internal structure of a source.
\end{abstract}

\section{Introduction}\label{sec:intro}

The process of measuring an image can, for many applications in
astronomy, be written as
\begin{equation}
D(\bfx)=(T*B)(\bfx)+N(\bfx).
\label{muckup}
\end{equation}
$\bfx$ represents a pixel position in a two-dimensional array,
or image, $D$ is the observed data, $T$ is the true image, $B$ is the
point-spread function (psf) of the measuring instrument, assumed to be 
invariant across the image, with $N$ being the associated statistical 
noise and $*$ represents the convolution operator such that
\begin{equation}
(f*g)(\bfx)=\int_n f(\bfy) g(\bfx-\bfy) \d\bfy.
\label{conv}
\end{equation}
$n$ is defined to be the number of pixels in the data image.
The task of an image reconstruction algorithm is to infer the true image given
the data, knowledge of the psf and the statistical properties of the noise.
For the simple case when there is no noise then, assuming that the
inverse of the beam exists, that there is no
structure on sub-pixel scales and ignoring complications
introduced by edge effects, the inferred truth, $\that$, can be
obtained using the convolution theorem and will exactly equal
$T$. Denoting the Fourier transform of $f(\bfx)$ by $\tilde{f}(\bfk)$,
$\that$ can be found by inverse transforming
\begin{equation}
\tilde{\that}(\bfk)=\tilde{D}(\bfk)/\tilde{B}(\bfk).
\label{deconv}
\end{equation}
The $n$ pieces of information in the data allow a perfect
reconstruction of the $n$ pixel values in the truth. With noise
switched on, the number of degrees of freedom in the solution increases
by $n$, while the number of constraints remains fixed, yielding an
ill-posed problem. Thus, the job of the reconstruction algorithm
becomes to decide which of the possible inferred truths is the best,
whatever that may mean.

An extension of the simple case described above involves filtering the
data to remove
the noise before inverse transforming equation (\ref{deconv}) to yield
$\that$. One particular, widely used example of this
is the Wiener filter (Wiener 1949; Rybicki \& Press 1992; Lahav \etal
1994; Bunn \etal 1994; Fisher \etal 1995). 
This procedure is non-iterative (it iterates to $\that=0$)
and the final inferred truth
is completely determined once guesses for the power spectra of the
true signal and the noise components have been made. 
If the assumed power spectra are correct then
the resulting filter will minimise over the whole image the
variance between the reconstructed and true signals. 
It can also be shown that the Wiener filter yields the
maximally probable inferred truth if the
true and noise values are normally distributed (Rybicki \&
Press 1992). For situations where these distributions are not
Gaussian, the optimal filter differs from the Wiener filter. 
While the Wiener filter method involves only a few transforms, and is
thus rapid, it would in general be desirable to break the degeneracy
in solution space with a technique that both does not require any
assumption about the nature of $T$ and produces optimal images for any true
signal distribution.

Another type of transform is the wavelet transform (\eg Slezak,
Bijaoui \& Mars 1990)
where the inferred truth is described using a set of basis functions
designed to extract information on a variety of scales. Once again,
this is a linear method that involves applying a transformation
to the data in order to extract information about particular scales of
interest.

An alternative approach to image reconstruction involves quantifying
the acceptability of a particular inferred truth, then iterating the
inferred truth until it becomes maximally acceptable.
One reasonable demand to make of $\that$ is that the resulting distribution
of residuals, defined as
\begin{equation}
R(\bfx)=D(\bfx)-(\that*B)(\bfx),
\label{resid}
\end{equation}
is statistically indistinguishable from that of the anticipated noise
$N$, \ie requiring a `good fit' to the data. The particular statistic
used to describe the extent of the misfit will depend on the nature of
$N$. For example, $\chi^2$ is a frequently employed statistic when noise values
are drawn from a normal distribution.

Having chosen a misfit statistic, approaches to reducing further the
acceptable solution space are more varied. A common and simple
route is to parametrise $\that$ and use the data to fit a small number
of parameters. This is very quick and effective, provided that the
prejudice contained in the parametrisation is appropriate. For
complicated images, a more sophisticated procedure is desirable.

To understand better where pixons fit into the story, consider
the following conditional probability equation with $M$ representing
all aspects of the model used to transform $\that$ to $D$:
\begin{equation}
p(\that,M|D)=\frac{p(D|\that,M)p(\that|M)p(M)}{p(D)}.
\label{condprob1}
\end{equation}
This is merely the probability of having a particular
combination of data, model and inferred truth, divided by the
probability of obtaining a particular data set. Since $D$ is measured
before it is used to infer $\that$, $p(D)$ is constant. In addition,
the desire to avoid introducing prejudice requires that $p(M)$ is
constant over all models. This leaves
\begin{equation}
p(\that,M|D) \propto p(D|\that,M)p(\that|M).
\label{condprob2}
\end{equation}
On the left hand side of this proportionality is the quantity that one
would like to maximise in the reconstruction, namely the probability
of a combination of inferred truth and model given the data. The first
term on the right
hand side is readily identified as the `likelihood', and the second term is
commonly called the `image prior'. From a Bayesian viewpoint, it
is reasonable to associate a probability to the plausibility of
obtaining a particular inferred truth once a model is specified,
despite the non-repeatable nature of the image prior.
This provides additional leverage
in the quest to reduce the acceptable solution space.
Note that even if the model is held fixed and one seeks
to maximise $p(\that|M,D)$ the above equation looks very similar
because in that case $p(M|D)=1$ and $p(\that,M|D)=p(\that|M,D) p(M|D)$.

In order to quantify the image prior, 
consider the situation of distributing $Z$ indistinguishable
photons randomly among $q$ buckets. Denoting the number of photons
in bucket $i$ by $z_i$, the probability of a particular distribution is
\begin{equation}
p(\{z_i\})=\frac{Z!}{q^Z \prod_{i=1}^q z_i!}.
\label{buckets}
\end{equation}
The most probable choice of $\{z_i\}$ can readily
be shown to be the one that has the same number of photons in each
bucket. This probability also increases when the number of buckets is
decreased; an Occam's razoresque tendency to favour simple descriptions. 
Referring back to the image prior, $p(\that|M)$ will be increased by
having the photons distributed evenly through the model buckets, and
by reducing the number of buckets.

This approach to image reconstruction was pioneered by Skilling (1989)
and Gull (1989). By analogy with statistical thermodynamics, they
related the maximisation of the image prior to a maximisation of the entropy
of the reconstructed truth. In the absence of additional information
which could shift the prior away from the default flatness, uniform
$\that$s are preferred over more complicated images that require a
less probable distribution of the indistinguishable photons in the
image pixels. 

These maximum entropy methods (MEM) are conventionally
applied in the pixel grid in which the data are measured.
However, if the real truth deviates from flatness then, to
the extent that the image prior is important relative to the likelihood term in
equation (\ref{condprob2}), this procedure will bias $\that$ away from
$T$. This suggests that the distribution of buckets into which the photons are
placed should be set according to the measured distribution $D$ if the
prior probability is to be truly maximised. Furthermore, one
implication from the above discussion is that the number
of buckets should be minimised in order to create the simplest, and
consequently most plausible, description for $\that$, rather than
using $n$ parameters because this is how many pixels there are in the
data image.
Considerations such as these led Pi\~na \& Puetter (1993, hereafter
PP93) to introduce the pixon concept. Unlike the uniformly arranged
pixels, pixons are able to adapt to the measured $D$ in order that the
information content is flat across the inferred truth when described
in the pixon basis. This adaptive `grid' essentially uses a higher
density of pixons to describe the inferred truth in regions where more signal
exists, and only a few very large independent pixons for the
background, or low signal-to-noise parts of an image.
Relative to MEM, the pixon method allows the
probability in equation (\ref{condprob2}) to maximise itself with
respect to one aspect of the model which conventional MEM keep fixed.
The task of maximising $p(\that,M|D)$ boils down to
finding the inferred truth containing the fewest pixons, which
nevertheless provides an acceptable fit to the data. Operationally,
the main difference between MEM and the pixon method is that MEM
explicitly quantifies the image prior and does a single maximisation
of $p(\that|M,D)$. In the pixon case this is split into successive
likelihood and image prior maximisations. Both MEM and the pixon
method require an assumption to be made concerning the noise
distribution in order that the goodness-of-fit can be quantified.

In summary, the pixon method is an iterative image reconstruction
technique that produces inferred truths that are smooth on a locally
defined pixon scale, and the iterations have a
well-determined finishing point, namely when the pixon distribution has
converged to the simplest state that yields an acceptable fit to the data.
This approach to image reconstruction has been applied to a
variety of astronomical data sets (\eg Smith \etal 1995; Metcalf \etal
1996; Dixon \etal 1996, 1997; Kn\"odlseder \etal 1996). A detailed
discussion of the theoretical basis of the pixon, as well as a list of 
applications of the method can be found in the paper by Puetter (1996). 

In the original pixon implementation described by PP93, the time taken
to perform a reconstruction scaled with the square of the total number
of pixels in the image. For typical astronomical images this meant
reconstructions would be impractically slow. 
While the discussion so far shows in principle the
advantages offered by the pixon approach, the implementation of the
idea remains to be specified. The main purpose of this
paper is to describe a speedy pixon algorithm that is limited by the
calculation of fast Fourier transforms, thus reducing the scaling of
the run time to $n {\rm log} n$ and allowing $256^2$ pixel images to be
reconstructed in a few minutes on a workstation. In
Section~\ref{sec:uses} the
method is applied to two simulated astronomical data sets. The results
are compared with a simple maximum likelihood method and the
robustness of the reconstructions is tested quantitatively using
Monte Carlo simulations. Puetter \& Yahil (1999)
have reported the existence of accelerated and quick pixon methods.

\section{Method}\label{sec:meth}

\begin{figure*}
\centerline{\epsfxsize=13cm \epsfbox{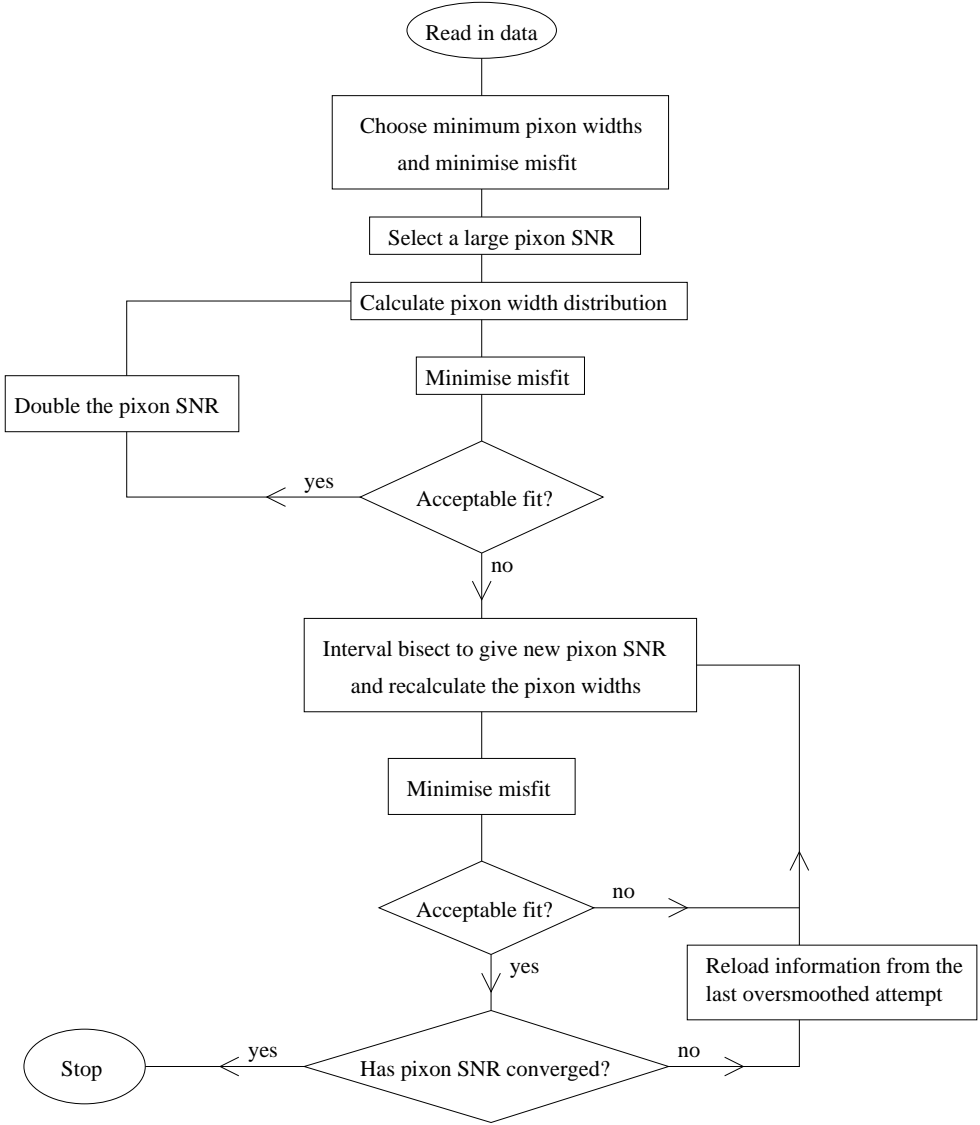}}
\caption{Flow diagram showing the structure of the algorithm. There is
an initial maximum likelihood type of fit, with all pixon widths
taking the minimum available value. The resulting pseudoimage is used
to find a pixon SNR that is too large to enable an acceptable fit to
be found. Then the largest pixon SNR giving rise to an acceptable residual
distribution is found by interval bisection.}
\label{fig:flow}
\end{figure*}

\subsection{Outline}\label{ssec:out}

The speedy pixon algorithm does not differ
greatly from that originally proposed by PP93. Namely, the
maximisation of the probability in equation (\ref{condprob2}) is
performed in an iterative fashion with each step consisting of a
change in the number of pixons being used to describe the inferred truth
followed by a conjugate gradient maximisation of the likelihood (or
equivalently minimisation of the misfit statistic) for the fixed
pixon distribution. These iterations are stopped when the inferred
truth is described using the fewest pixons that allow an adequate fit
to the data. The details of the pixon implementation are given in the
following section, before considering how they impact upon the
minimisation of the misfit statistic. A flow diagram outlining the
entire procedure is shown in Figure~\ref{fig:flow}.

\subsection{Specifics of the pixon implementation}\label{ssec:pixon}

For the examples described in Section~\ref{sec:uses}
the inferred truth is reconstructed in the same grid as the observed
data and thus contains $n$ values, one for each pixel. Given that the pixons
are to be less numerous than the pixels, these $n$ evaluations of the
inferred truth cannot be independent. In practice a
pseudoimage $H$ is defined in the pixel grid and $\that$ is
evaluated by correlating the pseudoimage values over a region with a
size determined by the local information content. This can be written
as
\begin{equation}
\that(\bfx)=\int_n P(\bfx,\bfy) H(\bfy) \d\bfy
\label{defi}
\end{equation}
where $P(\bfx,\bfy)$ is a weight function, the pixon shape,
defining how much signal the inferred truth at pixel $\bfx$ gathers
from the pseudoimage at $\bfy$. From this equation it can be seen that 
$\that(\bfx)$ merely represents the result of
convolving the pseudoimage with the pixon shape appropriate to pixel
$\bfx$. Thus, provided the number of distinct pixon shapes is finite
and independent of $n$, it is already apparent why the computational
time for this method is going to scale like $n {\rm log} n$ rather
than the direct summation $n^2$ of the original method.

The choice of pixon shapes and sizes will place constraints on
the types of truths that could possibly be recovered. Thus, using the
vocabulary adopted by Puetter (1996), it is important that the
richness in the pixon language is sufficient to enable a wide
variety of truths to be reconstructed. After all, it should be the
data which drives the reconstruction algorithm to an inferred truth,
rather than the algorithm knowing beforehand what it is going to see!
For the reconstructions
presented in Section~\ref{sec:uses}, a set of $n_{\rm pixon}$
circularly symmetric 2-dimensional Gaussians was used with widths,
$\{\delta_l: l=1,n_{\rm pixon}\}; \delta_{l+1}>\delta_l$, truncated at
$r=3\delta$. This pixon shape was found to give better
results than either an inverted paraboloid or a top hat for the
examples considered.

Constructing $\that$ in this way it was found that, for the
examples described in Section~\ref{ssec:xray} where there are
large ranges in signal strength across the image, the vast majority of
pseudoimage pixels were being set to zero during the reconstruction.
Consequently the $\that$ values in these pixels
were determined by the distance to the few pseudoimage pixels that
contained any signal and the appropriate pixon weight. To provide a little more
flexibility to the pseudoimage in such a situation, rather than relying
on the richness in the pixon set to enable a variety of images to be
inferred from a near delta function pseudoimage, the normalisation of
the $l$th pixon was chosen such that
\begin{equation}
\int_n P_l(\bfx,\bfy) \d \bfy =
\left(\frac{\delta_1}{\delta_{l}}\right)^\psi.
\label{pixbotch}
\end{equation}
With $\psi>0$, this means that in pixels having large pixon widths,
the inferred truth will be
suppressed relative to that calculated when all pixons are normalised
to the same value. The pseudoimage values in the low
signal regions, where the larger pixons are being used, are thus encouraged
to increase relative to those in the high signal parts of the image.
An unpleasant side effect of this is that the $\that$ values in high
signal pixels can become unduly influenced by the boosted regions of
the pseudoimage, so in calculating the signal, the following weights
were also included:
\begin{equation}
W(\delta(\bfx),\delta(\bfy)) = \left\{  \begin{array}{ll}
	1 & \mbox{if $\delta(\bfx) \geq \delta(\bfy)$} \\
	(\delta(\bfx)/\delta(\bfy))^\psi & \mbox{if $\delta(\bfx) < 
\delta(\bfy)$}
	\end{array}
	\right .
\label{pixweight}
\end{equation}
such that
\begin{equation}
\that(\bfx)=\int_n P(\bfx,\bfy)
W(\delta(\bfx),\delta(\bfy)) H(\bfy) \d\bfy.
\label{newsig}
\end{equation}
This down-weights the contribution to $\that$
gathered from the pseudoimage pixels with larger $\delta$s. Values of
$\psi$ between $0$ and $1$ were used for the reconstructions in
Section~\ref{sec:uses}. The introduction of this extra weight means
that the calculation of $\that$ no longer involves just a single
convolution. However, by splitting the integral into $n_{\rm pixon}$
separate convolutions of $P_l$ with a $W(\delta_l,\delta(\bfy))$-weighted 
$H$, the $n \log n$ scaling can be preserved.

The procedure for defining the distribution of pixon widths is based
on the desire to have the same amount of information in each
pixon. Information content can be parametrised through a pixon
signal-to-noise ratio (SNR). This is defined as
\begin{equation}
S(\delta(\bfx))=\frac{\delta(\bfx)\int_n
P(\bfx,\bfy) \that(\bfy) \d\bfy}{\sqrt{\int_n
P(\bfx,\bfy) \sigma^2(\bfy) \d\bfy}} 
\left(\frac{\delta(\bfx)}{\delta_1}\right)^{\frac{\psi}{2}}.
\label{pixsnr}
\end{equation}
$\sigma(\bfy)$ is the anticipated amplitude of the noise in pixel
$\bfy$. The final term just represents the removal of the
non-normalisation of the pixons, and the first $\delta(\bfx)$
is there because the mean SNR within the pixon is being calculated.
For a specified pixon SNR, the pixon width distribution is chosen such that
all pixels have the minimum available $\delta$ that provides at least
the required SNR. 

For the reconstructions described in
Section~\ref{ssec:xray}, it was found that relaxing the need for the
inferred truth to have a constant SNR in each pixon led to 
superior results. With a flat inferred truth in the pixon basis,
the higher signal regions of the data image received larger, and thus less
likely, reduced residuals. In order to decrease this discrepancy, the
required pixon SNR was decreased in the high signal pixels.
The pixon SNR appropriate for pixel $\bfx$ was defined to be
$f_{\rm SNR}(\bfx)$ times the global value, where
\begin{equation}
f_{\rm SNR}(\bfx)=\max(\upsilon,\sqrt{{\rm max}(0.,1-(1-\upsilon^2)r(\bfx))}),
\label{snrbotch}
\end{equation}
with $r(\bfx)$ being the convolution of $|R(\bfx)/\sigma(\bfx)|$
with the correctly normalised pixon appropriate to each pixel,
and $\upsilon$ a constant chosen to lie in the range $[0,1]$. While
this is a very ad hoc addition to the method, it encapsulates the
desire to provide more freedom in the badly fitting regions of the
reconstructed image, rather than keeping rigorously to the maximum
entropy flat $\that$ solution.

The remaining part of the pixon story, is to describe the fashion in
which the pixon SNR is iterated. To define the first $\that$, in order
that the SNR can be computed via equation (\ref{pixsnr}), a maximum
likelihood type fit is performed using $\delta(\bfx)=\delta_1 ~\forall \bfx$
and starting from a flat $\that$ with the mean value of $D$. An
initial pixon SNR is chosen such that the resulting pixon
distribution has too few degrees of freedom to allow a good fit to be
obtained. Interval bisection is then used to find the maximum
acceptable pixon SNR. At each stage, the last over-smoothed (\ie badly
fitting) $\that$ is used to infer the new pixon distribution. This
decreases the chance of introducing spurious structures into the
reconstruction. In
practice, once the pixon SNR converges to about $20$ per cent,
$\that$ is insensitive to further refinement and the procedure is stopped.

\subsection{Specifics of the likelihood calculation}\label{ssec:like}

The other half of the reconstruction method involves the likelihood term, and
the procedure used to maximise this for a fixed pixon distribution. 
A likely $\that$ is one with a small misfit statistic, ie one that
`fits the data well'.
Gaussian and Poisson distributed noise are relevant for the examples in
Sections~\ref{ssec:sz} and~\ref{ssec:xray},
so reconstructions were attempted using the $\chi^2$ and
\begin{equation}
\chi^2_{\gamma}=\sum_{i=1}^{n}\frac{(R(i)+{\rm min}(D(i),1))^2}{D(i)+1}
\label{chi2g}
\end{equation}
(Mighell 1999) misfit statistics respectively. 
However, superior results were obtained by
employing the $E_R$ statistic of Pi\~na \& Puetter (1992, PP92). Rather
than just taking into account the amplitudes of the residuals, this measures
their spatial autocorrelation function. 
In more detail, $E_R$ is defined as
\begin{equation}
E_R=\frac{1}{n}\sum_{\bfz} A_R(\bfz)^2,
\label{erdef}
\end{equation}
where $\bfz$ represents a 2-dimensional lag in pixel space and $A_R$
is the autocorrelation of the reduced residuals
\begin{eqnarray}
A_R(\bfz) &=& (R/\sigma \otimes R/\sigma)(\bfz) \cr
&=& \sum_n (R/\sigma)(\bfy+\bfz) (R/\sigma)(\bfy).
\label{ardef}
\end{eqnarray}
As PP92 showed, the expected value of $E_R$ is equal to the number of
lags included in the summation in equation (\ref{erdef}), and the extent over which lag terms are useful is determined by
the size of the instrumental psf. For the applications described below,
an acceptable fit is defined to be one that has $E_R < 3~+$~the number of lag
terms being considered. This is true $\sim 90$ per cent of
the time for normally distributed noise or Poisson distributed noise
when the mean signal is $\gsim 1/\sqrt{n}$. 

To minimise the misfit statistic, 
the Polak-Ribiere conjugate gradient algorithm in Numerical Recipes (Press
\etal 1992) was used. Some alterations were made to tailor the general
purpose routine to this specific task, as detailed in
Appendix~\ref{app:cjgd}.
In order to have non-negative inferred truths, PP92 suggested using
transformed pseudoimage values, $H_\t$, in the minimisation where
\begin{equation}
H(\bfx)=\alpha(H_\t(\bfx)+\sqrt{H_\t(\bfx)^2+\beta}),
\label{vartrans}
\end{equation}
with $\alpha=0.5$ and $\beta=1$. This transformation was used for the
example in Section~\ref{ssec:sz}, but created a poor reconstruction of
the low-signal regions in the examples in Section~\ref{ssec:xray}. For
these cases, setting $H_\t=H$ and simply truncating negative values
yielded superior results without significantly affecting the speed of
the code.

The calculation of the gradient of the misfit statistic with respect to the
transformed pseudoimage values is a bit messy. After all, changing
$H_\t$, or effectively $H$, alters the inferred truth in surrounding
pixels. This is then convolved with the instrumental psf before the
residuals and hence the misfit are calculated. Appendix~\ref{app:dmisf}
contains the results of this calculation, but the important point is
that it can be split up into correlations and convolutions, thus
maintaining the $n \log n$ scaling of the algorithm.

\section{Applications}\label{sec:uses}

The speedy pixon algorithm was applied to two simulated
astronomical data sets. In the first case, the challenge of
identifying galaxy clusters in Cosmic Microwave Background (CMB) maps
was considered for an instrument with specifications like those of the
Planck surveyor. The formalism described above will be extended to
deal with the multifrequency and multicomponent natures of the data
and truth respectively. In the second example some simulated
`$\beta$'-profiles convolved with a large psf, such as the ASCA 
X-ray detector would measure when pointed at galaxy clusters, 
are reconstructed.

\subsection{Multiwavelength cluster detection in simulated CMB
data}\label{ssec:sz}

\begin{table*}
\begin{center}
\caption{Lists of
(1) the central Planck observing frequencies employed here (GHz);
(2) the Gaussian full width half maxima (arcmin); 
(3) the $1\sigma$ Gaussian noise per $1.5$ arcmin square pixel in mJy
(for $14$ months of observations); 
(4) the conversion factor relating thermodynamic CMB $\Delta T/T$, in
units of $10^{-6}$, to the change in flux in mJy in a pixel; 
(5) the conversion factor relating the Comptonisation parameter $y$,
in units of $10^{-6}$, to the change in flux in mJy in a pixel; 
(6) the rms cluster $T*B$ per pixel (mJy); 
(7) the maximum amplitude of cluster $T*B$ per pixel (mJy); 
(8) the rms intrinsic CMB $T*B$ per pixel (mJy); 
(9) the maximum amplitude of intrinsic CMB $T*B$ per pixel (mJy).}
\begin{tabular}{lllllllll} \hline
~~~~~(1) & ~(2) & ~~(3) & ~(4) & ~~~(5) & ~~~~(6) & ~~~~~~(7) &
~~~~~~(8) & ~~~~~~(9) \\
Frequency &
fwhm &
Noise &
$\omega_{\rm SZ}(\j)$ &
$\omega_{\rm CMB}(\j)$ &
rms $T_{\rm SZ}*B$ &
max $|T_{\rm SZ}*B|$ &
rms $T_{\rm CMB}*B$ &
max $|T_{\rm CMB}*B|$ \\
~~~~100 & ~10.7 & ~1.2 & -0.19 & ~~0.124 & ~~0.20 & ~~~~~~5.1 & ~~~~~~4.6 & ~~~~~~17.4 \\
~~~~143 & ~8.0 & ~2.2 & -0.20 & ~~0.197 & ~~0.23 & ~~~~~~7.8 & ~~~~~~7.6 & ~~~~~~29.1 \\
~~~~217 & ~5.5 & ~3.1 & -1.9$\times 10^{-3}$ & ~~0.249 & ~~2.3$\times 10^{-3}$ & ~~~~~~0.11 & ~~~~~~9.9 & ~~~~~~38.5 \\
~~~~353 & ~5.0 & ~5.8 & 0.34 & ~~0.155 & ~~0.43 & ~~~~~~20.6 & ~~~~~~6.2 & ~~~~~~24.1 \\
\hline
\end{tabular}
\end{center}
\label{tab:planck}
\end{table*}

The Planck surveyor satellite (Tauber, Pace \& Volont\'e 1994) is expected to
return maps of the sky in a number of microwave wavelength ranges. In
addition to the intrinsic CMB fluctuations, a number of interesting foregrounds
will also contribute to these maps. One such contribution will come
from the Sunyaev-Zel'dovich (SZ) effect produced when CMB photons are
inverse Compton scattered during their passage through the ionised gas
in galaxy clusters (Sunyaev \& Zel'dovich 1972). The distinctive
spectral distortion created by the net heating of CMB photons in the
directions of galaxy clusters should enable some thousands of clusters
to be detected by Planck (Hobson \etal 1998).

\subsubsection{Additional formalism}\label{sssec:cmbform}

The formalism described here will assume that the spectral dependence
of each of the components (\ie CMB, SZ and noise) is known and 
constant over the region
being observed, although in principle this could also be left as a
part of the model over which the probability is maximised. For each of the
$n_c$ components, the spectral template will be denoted by
$\omega_\c(\j):\c=1,n_c;\j=1,n_\lambda$, where $n_\lambda$ is the
number of wavebands in the observed data. Thus the inferred truth in
waveband j can be written
\begin{equation}
\that(\bfx,\j)=\sum_{\c=1}^{n_c} \omega_\c(\j) \hspace{-0.1cm} \int_n
\hspace{-0.1cm} P_\c(\bfx,\bfy) W(\delta_\c(\bfx),\delta_\c(\bfy)) 
H_\c(\bfy) \d\bfy.
\label{multic}
\end{equation}
Each component has its own pixon distribution and pseudoimage denoted
by the c subscript. The
pseudoimage variables for the intrinsic CMB and thermal SZ components
are the thermodynamic $\Delta T/T$ and the Comptonisation parameter $y$
respectively, both in units of $10^{-6}$. 
Note that $\Delta T/T$ can take positive or negative
values so the transformed pseudoimage is chosen to equal the
pseudoimage for this component, whereas equation (\ref{vartrans}) is used to
ensure that $H_{\rm SZ}(\bfx) (\equiv y)$ is non-negative in all pixels.
For the reconstructions presented below, the value of
$\psi$, as defined in equation (\ref{pixbotch}), was set to zero such that all
pixons were normalised to unity. In practice, it may be beneficial to
allow $\psi$ to be a function of component.

The multiwavelength data, and the fact that the CMB component
can take positive or negative values, require a
definition of pixon SNR differing from that in
equation (\ref{pixsnr}). This is chosen to be
\begin{equation}
S(\delta_\c(\bfx))=\frac{\delta_\c(\bfx) \int_n
P_\c(\bfx,\bfy) |\that_\c(\bfy,1)| \d\bfy}{\sqrt{\int_n
P_\c(\bfx,\bfy) \sigma^2(\bfy,1) \d\bfy}}
\left(\frac{\delta_\c(\bfx)}{\delta_{1,\c}}
\right)^{\frac{\psi}{2}}.
\label{pixsnr2}
\end{equation}
The $1$ index with the $\that$ and $\sigma^2$ terms represents that
only the first waveband is being used to define the signal-to-noise
ratio and $\delta_{1,\c}$ is the smallest pixon width for component $\c$.
For the type of data simulated here, with the intrinsic CMB component
dominating the thermal SZ component, it is possible, if the required
pixon SNR is the same for all components, to find an
acceptable fit to the data without the need for a cluster component in
the inferred truth. In order to allow a better reconstruction of the
weaker component (\ie a reconstruction that includes some signal),
it is necessary to introduce 
factors $g_{\rm SNR}(\c)$ such that the actual pixon SNR requested for pixons
representing component $\c$ is $g_{\rm SNR}(\c)$ times the default value.
While these factors could be left for the pixon algorithm to evaluate
in a Bayesian fashion, this would be rather time consuming. In practice
the relative strengths of the components were set to 
$g_{\rm SNR}=1$ and $0.1$ for the intrinsic CMB and thermal SZ components
respectively in order that clusters were found, without introducing
many spurious sources. 

\begin{figure*}
\centerline{\epsfxsize=18cm \epsfbox{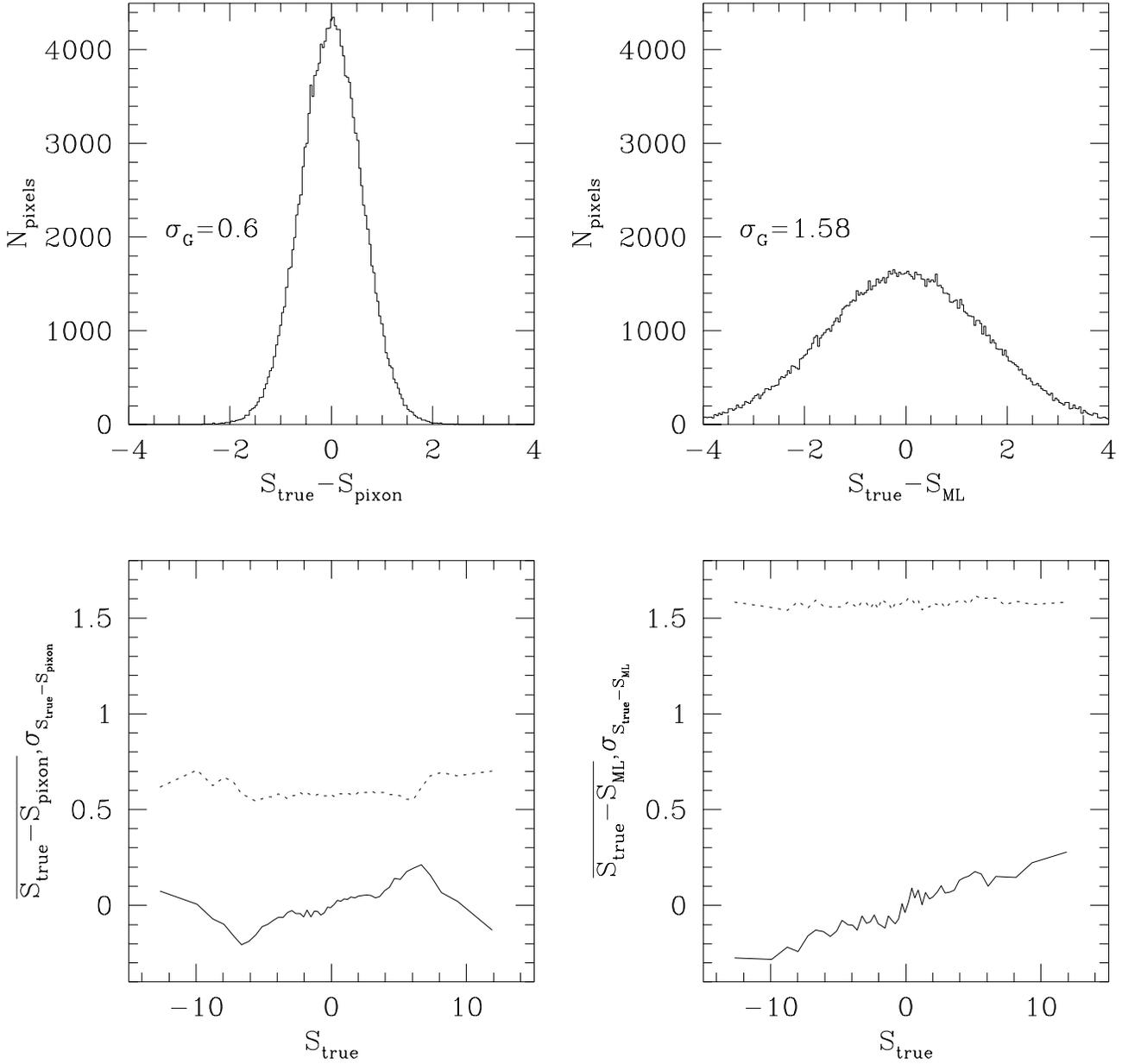}}
\caption{The top left panel contains a histogram of the 
difference between the true and pixon-inferred intrinsic CMB component
fluxes at $100$ GHz in
each pixel. Flux units are mJy for all panels in this figure. The
width of the best-fitting Gaussian is also given. In the lower left
panel, the solid line represents the average difference between the
true and pixon-inferred intrinsic CMB component fluxes as a function
of the true pixel flux, and the dashed line traces the standard deviation of
the reconstruction error. The two right hand panels show the
corresponding results for the ML reconstruction. In both cases, the
Gaussian fits to the reconstruction errors are not shown because they
essentially lie on top of the histograms.}
\label{fig:cmbcomp}
\end{figure*}

\begin{figure}
\centerline{\epsfxsize=7.8cm \epsfbox{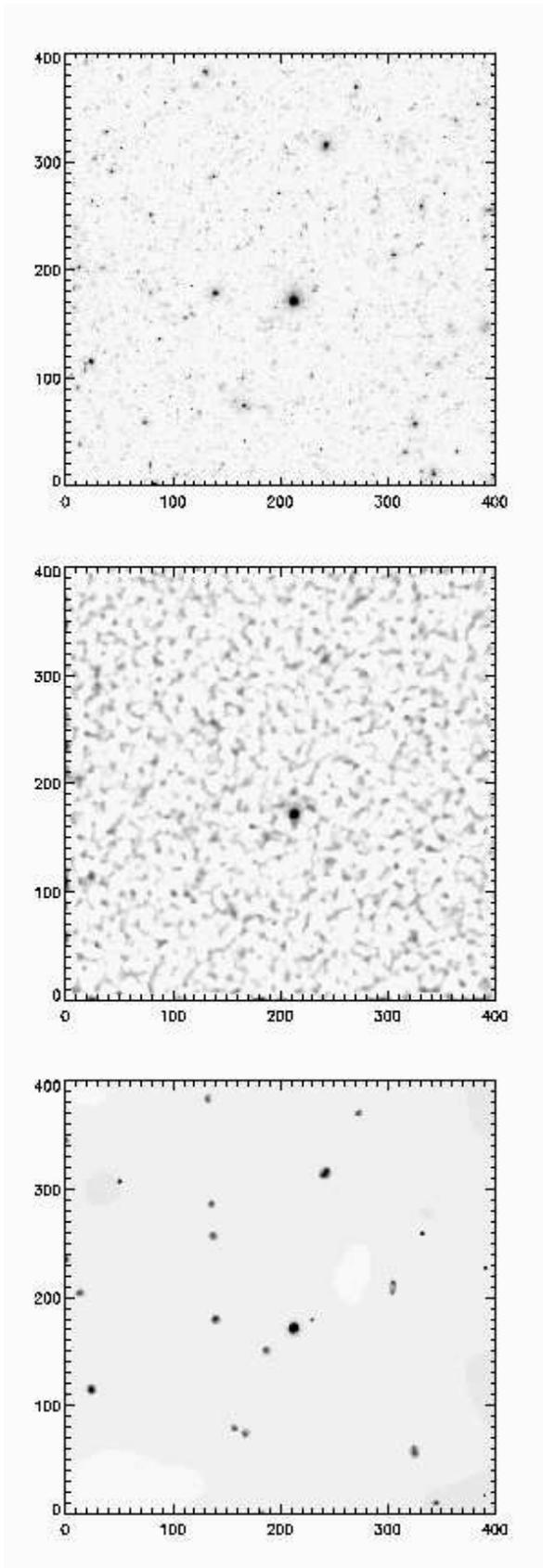}}
\caption{The true thermal SZ $y$ map is shown in the
top panel, and the ML and pixon-inferred truths are contained in the
second and third panels respectively. Axes are labelled in pixels.}
\label{fig:itall}
\end{figure}

The $\upsilon$ parameter for determining the variation in pixon SNR
across the image for a given component was set to $1$ for both the CMB and SZ
components. This enforced a flat pixon SNR across each of the component
maps ($f_{\rm SNR}=1$). If some SNR variation was to be used, then an average 
over wavebands of the reduced residuals convolved with the local pixon
shape would need to be calculated, rather than the monochromatic version 
contained in equation (\ref{snrbotch}).

To deal with the sharp edge in the observed data, the reconstructed images
were allowed to extend beyond the original data pixel grid. A total of
$512^2$ pixels were used and those lying outside the input data image had
$R(\bfx)=0$ defined in them, but were otherwise treated identically
with the rest of the reconstructed image pixels. This buffer region
allows the pixon SNR to be sensibly defined, so that the inferred
truths have the same sensitivity across all of the observed image
while not directly influencing the calculation of the misfit statistic.

The computation of the misfit statistic includes $n_\lambda$
different residual maps and the definition of an acceptable value is
modified accordingly. In addition, the misfit minimisation is
performed over $n \times n_c$ variables.

\subsubsection{Data production}\label{sssec:szdat}

A $400^2$ pixel, $10$ degree square field of simulated CMB sky, was created
including both intrinsic CMB fluctuations and thermal SZ distortions
produced by clusters,
The intrinsic CMB map was a realisation of the standard Cold Dark
Matter model using the power spectrum returned by CMBFAST (Seljak \&
Zaldarriaga 1996). The thermal SZ map was produced by creating some
templates from
the hydrodynamical galaxy cluster simulations of Eke, Navarro \& Frenk
(1998) and then pasting these, suitably scaled, at random angular
positions with mass and redshift distributions according to the
Press-Schechter formalism (Press \& Schechter 1974).
The thermal SZ Comptonization parameter, $y$, and the thermodynamic
$\Delta T/T$ of the intrinsic CMB fluctuations were converted to fluxes
per pixel in mJy\footnote{$1$ mJy $\equiv 10^{-29}$ W m$^{-2}$ Hz$^{-1}$}
in each of four wavebands by multiplying by
$\omega_\c(\j)$ as listed in columns 4 and 5 of Table 1.
These four combined truths were then `observed' by applying the
relevant Gaussian beam and adding the pixel noise appropriate to each
wavelength (see columns 2 and 3 of Table 1).

The final four columns in Table 1 show the maximum $|T_\c*B|$ and
rms $T_\c*B$
per pixel produced by each of the two components separately. These
numbers demonstrate both the high SNR with which Planck
will observe the intrinsic CMB fluctuations and the relatively low
SNR of even the brightest cluster in the field,
after convolution with the instrumental psfs.

\subsubsection{Results}\label{sssec:szres}

The sets of pixon widths for the two components were chosen to be
$3\delta_{\rm CMB}(\bfx)=5, 7, 9$ and $3\delta_{\rm SZ}(\bfx)=5, 9,
80$. For the intrinsic CMB, the signal-to-noise ratio is so good that
the pixons do not need to be very large before they prevent an
acceptable fit from being found. In the SZ clusters case, 
the selection of a small and a very large pixon size allows the 
reconstruction of sharp features in a smooth background. The
intermediate-sized pixon helps the pixons bridge the gap between
background and cluster as the pixon SNR is reduced and the map becomes
progressively less correlated. These SZ pixon widths are selected to match
the scales of the features expected to be present for this component map.
About $20$ minutes of computer time were required to perform the
multicomponent and multiwavelength speedy pixon
reconstruction of the $400^2$ pixel image (padded up to $512^2$ pixels).
For comparison a maximum likelihood (ML) reconstruction with all pixons 
being set to be pixel-sized was also performed. This took approximately $3$ 
minutes to complete. 

In Figure~\ref{fig:cmbcomp}, the pixon and ML-inferred truths for the
intrinsic CMB flux at $100$ GHz are compared with the true values in
each pixel.
The top panels show the distributions of reconstruction errors per
pixel for the two methods, along with the widths of the best-fitting
Gaussians for these distributions. Comparing the raw data (\ie
including SZ clusters, convolution with the beam and noise) with the
true intrinsic CMB pixel values, leads to a best-fitting Gaussian
width of $1.70$ mJy, so it is apparent that both pixon and ML
reconstructions have cleaned the data to some extent, although the
narrowing of the error distribution is significantly better for the
pixon case.
The lower panels show trends for both the mean (solid lines) and standard
deviation (dotted lines) of the flux errors as a function of the true
pixel flux. In the ML case, the scatter in the flux errors is large
and approximately independent of the true signal, whereas for the pixon
reconstruction the scatter is suppressed but increases when the signal
is strong and the pixon width being used becomes smaller. Where the
scatter in the error increases, the mean difference between the pixon-inferred
and true signals decreases. This shows that for pixels with absolute
values of intrinsic CMB $100$ GHz flux exceeding $\sim 6$ mJy, a
smaller	 pixon width has been selected and the fit has improved. The
choice of pixon widths is thus very important in determining these
results. In the ML case, the trend in the mean error shows that the
peak sizes are systematically underestimated.

Figure~\ref{fig:itall} is a greyscale comparison of the true (top
panel), ML-inferred (middle) and pixon-inferred (bottom) cluster y
maps. It is very apparent that the pixon reconstruction has greatly
suppressed the noise relative to the ML effort. There are a few
sources in the pixon reconstruction that do not correspond with single
identifiable sources in the actual truth. In regions where the density
of small clusters is particularly high, the pixon algorithm has a
tendency to place a single bright source to model the
emission. However, relative to the ML effort, the compression
of the reconstructed information is very clear.
The pixon algorithm has essentially already made the decision as to
which of the many ML sources are statistically justifiable.
Reducing the SZ pixon SNR relative to that of the intrinsic CMB using
the $g_{\rm SNR}(SZ)$ parameter would allow the pixon algorithm to detect
clusters with smaller fluxes, albeit with an increased risk of
producing spurious sources.
The mean Compton $y$ parameters per pixel in units of $10^{-6}$ are
$0.80$, $1.25$ and $0.79$ for the true, ML and pixon images
respectively, so the pixon algorithm does a good job of conserving
the entire thermal SZ flux, in contrast to the ML technique.

As an aside, the inclusion of an intrinsic CMB component does not
affect the cluster detection efficiency significantly. The
important quantity is the signal-to-noise ratio with which the
clusters alone would be observed. Other foregrounds such as dust and
Galactic free-free and synchrotron emissions are unlikely to vary on
small scales and should not greatly affect the ability of the
algorithm to detect clusters (see \eg Hobson \etal 1998).

\subsection{Simulated ASCA X-ray cluster data with Poisson distributed noise}\label{ssec:xray}

\begin{figure*}
\centerline{\epsfxsize=18cm \epsfbox{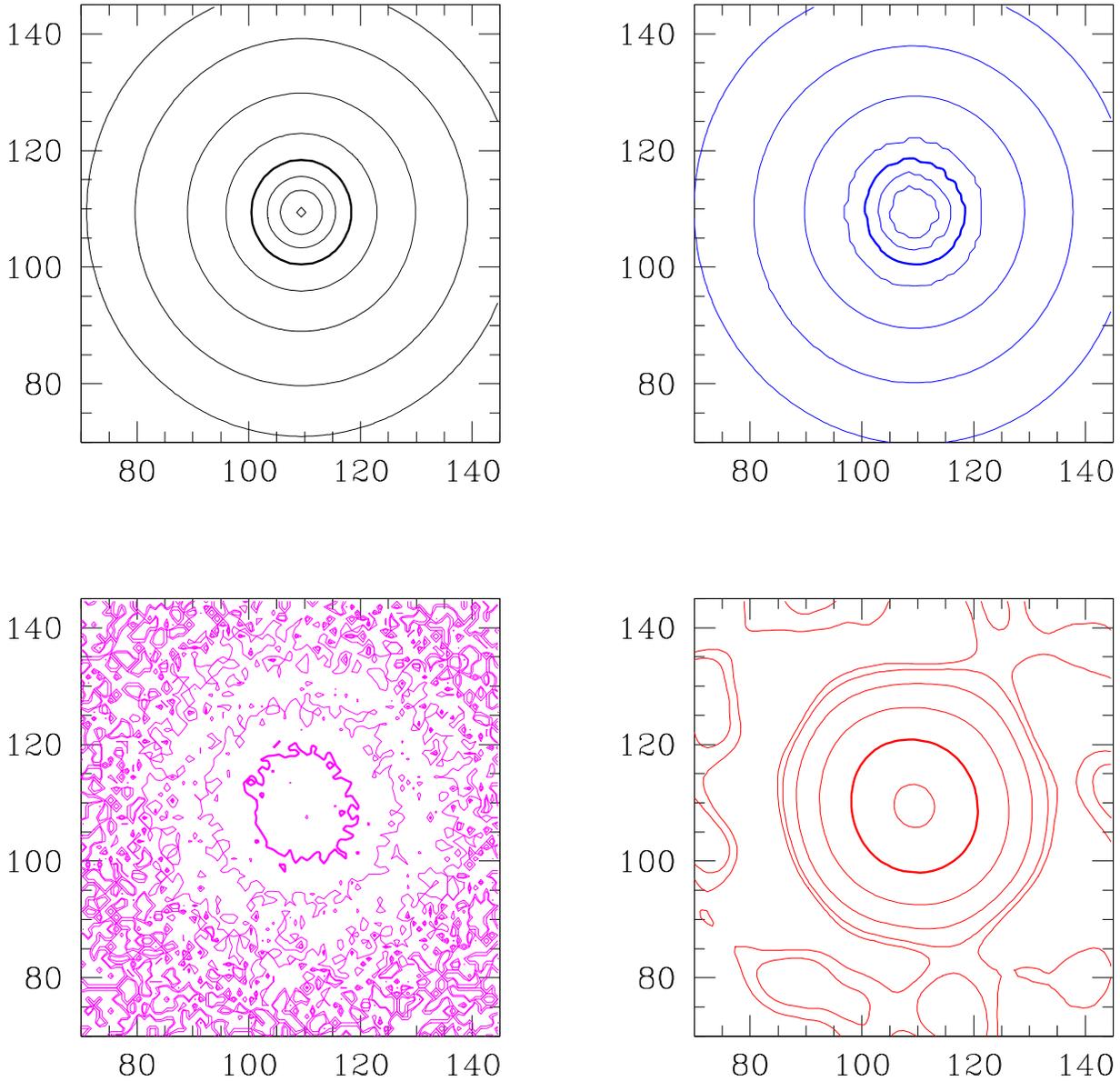}}
\caption{Contour plots for the X-ray cluster example in
Section~\ref{ssec:xray} showing the central region of the high signal-to-noise
true $\beta$ profile (top left), one noisy realisation of it (lower left), a
pixon-inferred truth (top right) and the ML-inferred truth (lower
right). The contour levels are $0.5, 1, 3, 10, 30$ (bold),$ 70, 150$ and
$300$ counts per pixel and the scales on the axes are numbers of pixels.}
\label{fig:manyim}
\end{figure*}

X-ray imaging of galaxy clusters using the ASCA satellite involves
convolution with a broad energy-dependent instrumental psf.
The fact that the psf varies with position in the
image will be neglected in the following examples, because
equation (\ref{muckup}) is inapplicable in such situations and, if the
bulk of the emission is very concentrated then the
instrumental psf will be approximately constant over the region of 
interest, in which case this treatment would be valid. ASCA has a
particularly broad psf for an X-ray instrument, so its use for imaging
might appear rather surprising. However, the good spectral resolution,
coupled with the energy-dependence of the psf creates a situation
where an image reconstruction algorithm could, for instance, be
usefully employed in
determining temperature maps of the ionised gas in X-ray clusters.
Poisson, rather than Gaussian, noise is appropriate for these images
where the numbers of photons per pixel is small.

\subsubsection{Data production}\label{sssec:xraydat}

Two different surface brightness profiles were created in a $256^2$
pixel grid according to 
\begin{equation}
\Sigma(r)=\frac{\Sigma_0}{(1+(r/r_c)^2)^{3\beta/2}}+b.
\label{betaprof}
\end{equation}
This is the $\beta$-profile proposed by Cavaliere \& Fusco-Femiano
(1976) to represent cluster X-ray surface brightness profiles with $b$
corresponding to an additional background contribution. The high
signal-to-noise model had $(\Sigma_0,r_c,\beta,b)=(320~{\rm
counts/pixel}, 5~{\rm pixels}, 0.7, 0.1~{\rm counts/pixel})$ whereas
the low signal-to-noise truth used $(7, 6, 0.7, 0.001)$. These truths
were chosen to be similar to what the ASCA satellite would have
seen when looking at a cluster in the energy ranges $1-2$~keV and $7-8$~keV.
A non circularly symmetric ASCA-like psf having a full width half
maximum of $\sim 10$ pixels was applied to these two
truths and $10$ Monte Carlo realisations of the resulting $T*B$s were made.
After smearing out with the beam, the maximum counts per pixel were
$\sim 70$ and $1.5$ for the two data sets, giving signal-to-noise
ratios of $\sqrt{70}$ and $\sqrt{1.5}$ at the peak of the emission.

\subsubsection{Results}\label{sssec:xrayres}

Reconstructions were performed using $12$ pixons
ranging in size from $\delta_1\approx 1$ pixel to $3\delta_{n_{\rm
pixon}}\approx 100$ (separated by factors of $\sim 1.3$), 
$\psi=0.8$ and $\upsilon=0.6$.
These choices were kept the same for both the data sets. For the
Poisson distributed noise the expected amplitude of the
noise in pixel $\bfx$ was set such that $\sigma(\bfx)^2=(\that*B)(\bfx)$.
The comparison ML reconstructions were performed by setting all pixon
widths across the pseudoimage to equal one pixel.

For the low signal-to-noise example, a tolerable fit according to
$\chi^2_\gamma$, could actually be obtained with a flat $\that$. Only
when the misfit statistic was changed to $E_R$ was it
necessary to insert a source into the inferred truth in order to
produce a good fit. That is, the correlated residuals produced when a uniform
$\that$ was used to describe the weak source were sufficiently small
that their amplitudes were statistically acceptable. However the spatial
correlation of the residuals did have the power to discriminate between
this residual field and the anticipated noise. 

Figure~\ref{fig:manyim} shows the central regions of the `X-ray
cluster' for the high signal-to-noise data. 
The top-left panel shows the truth, and one of the
realisations of the observed data is shown beneath this. Both the
smearing out of the sharply peaked emission and the introduction of
noise are very evident. The pixon-inferred truth for this particular
$D$ is shown in the top-right panel and the corresponding ML-inferred truth is
contained in the final panel. It can be seen that the noise is
greatly suppressed by the pixon method and much of the peaked emission
has been recovered on sub-psf scales. The ML reconstruction also
removes noise in the central regions, but at large radii spurious
features are introduced, essentially fitting to the noise. Also, at
small radii, the ML deconvolution does not do a good job of recovering
the unsmeared profile.

\begin{figure}
\centerline{\epsfxsize=11cm \epsfbox{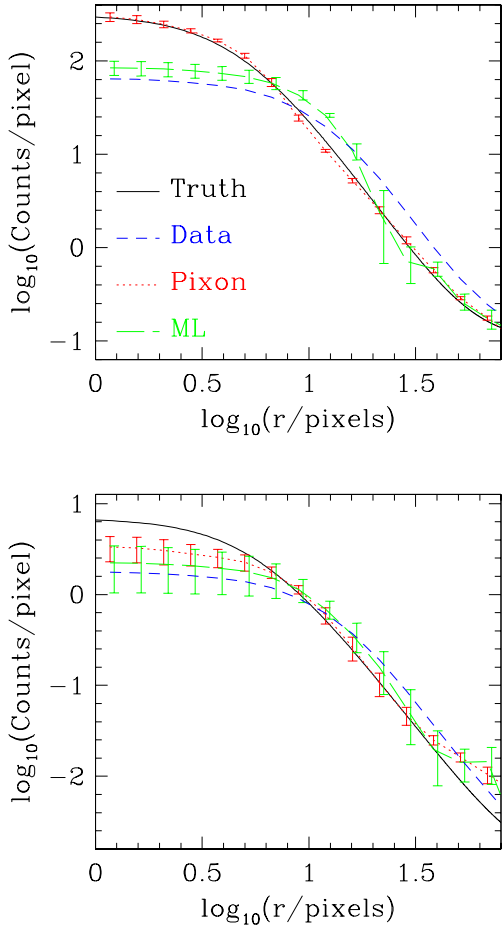}}
\caption{Azimuthally averaged profiles, showing the performance of the
pixon (dotted lines) and ML (long-dashed lines) algorithms for
reconstructing high (upper panel) and low signal-to-noise beta
profiles. Error bars on the reconstructed results represent the standard
deviation of $10$ Monte Carlo realisations of the same truth. For
clarity, the error bars for the ML results have been displaced 0.02 to
the right. The full width half maximum of the beam is $\sim 10$
pixels. Solid and short-dashed lines represent the true $\beta$
profiles and the data respectively.}
\label{fig:rprof2}
\end{figure}

More detailed results concerning the radial surface
brightness profiles are shown in Figure~\ref{fig:rprof2}, for both the
high and low signal-to-noise data sets. The mean pixon-inferred
profiles are drawn as dotted lines along with error bars showing the standard
deviation of the individual Monte Carlo reconstructions. Long dashed
lines represent the corresponding ML quantities. It is
apparent and reassuring, particularly for the high signal-to-noise
data, that the pixon algorithm tends to produce similar profiles,
independent of what noise realisation has been used. The
ML results show a significantly larger dispersion
in the reconstructions arising from the different noise realisations.
However, for the pixon reconstructions, the
error bars are sufficiently small that the systematic deviations
between the inferred and actual truths can be seen. These are produced
by the lack of richness in the pixon description which leads to a preference 
for particular
profiles. Nevertheless it is encouraging that the speedy pixon algorithm can
yield such results for two very different quality data sets.
In addition, the suppression of noise is very effective, with no hint
of any spurious sources being produced in any of the realisations
even for the low signal-to-noise simulations, in contrast to the ML
reconstructions.

\section{Conclusions}\label{sec:conc}

The details of a speedy pixon method for image reconstruction have
been given. This algorithm reduces the run time from the $n^2$ in the
original procedure described by PP93 to $n \log n$, making the
treatment of $256^2$ pixel images possible using only a few minutes on a
typical workstation. The application of the method to two types of
simulated data sets shows its ability to detect sources
in low signal-to-noise data without introducing spurious objects, as
well as deconvolving the instrumental psf. These results are a marked
improvement over a simple maximum likelihood reconstruction procedure
which is applied in the data pixel grid and includes a uniform image
prior term. A more detailed study of the ability of the pixon method to
find clusters through their S-Z distortion of CMB maps is in progress,
including a comparison with MEM.

\section*{ACKNOWLEDGMENTS}

I would like to thank George Efstathiou for his helpful comments,
including pointing out to me the existence of pixons, R\"udiger
Kneissl for his CMB map and many
useful discussions, David White for providing ASCA details and helpful
suggestions, Shaun Cole for FFT assistance and Doug Burke, Ofer Lahav
and Radek Stompor for general enlightenment.
This work was carried out with the support of a PPARC postdoctoral fellowship.

\begin{appendix}
\section[]{Conjugate gradient minimisation details}\label{app:cjgd}

The vast majority of the run time of the speedy pixon method is spent 
calculating the derivative of the misfit statistic with respect to the
transformed pseudoimage values, $H_\t$, and evaluating the inferred
truth for a given $H_\t$. A couple of simple changes to the
Numerical Recipes line minimisation routine, linmin, significantly
reduce the number of function and derivative calls, and thus merit a
mention here. (A more detailed discussion of some of these issues is
contained at http://wol.ra.phy.cam.ac.uk/mackay/c/macopt.html.)
Firstly, the default tolerance requested by linmin is $\sim 10^3$
times more stringent than need be for this application. Secondly, the
initial guesses at bracketing the step size required to reach the
function minimum
along the chosen direction can be made more efficiently. Rather than
keeping them fixed at $0$ and $1$, these sizes should reflect the fact that
the different steps in parameter space are likely to have similar 
magnitudes. Thus
the initially guessed step size should be inversely proportional to
the modulus of the vector along which the step is to be taken.
In addition, allowing these estimates to adapt to
previous values also leads to a more rapid minimisation. Tuning
the routine along these lines leads to a speed up of about an order of
magnitude for the examples considered in this paper.

\section[]{Calculation of the misfit statistic derivatives}\label{app:dmisf}

The conjugate gradient misfit statistic minimisation requires the
evaluation of the partial derivatives of the statistic (hereafter
labelled $\mu$) with respect to the
transformed pseudoimage variables. The chain rule for differentiation
gives
\begin{equation}
\frac{\partial{\mu}}{\partial H_\t(\bfx)} = \frac{\partial{\mu}}
{\partial H(\bfx)} \frac{\partial H(\bfx)}{\partial H_\t(\bfx)}.
\label{appeq:chain}
\end{equation}
Defining the partial derivative of $\mu$ with respect to
$(\that*B)(\bfx)$ by $F(\bfx)$, the derivative of $\mu$ with respect
to the pseudoimage can be written as
\begin{equation}
\frac{\partial{\mu}}{\partial H(\bfx)}=\sum_{l=1}^{n_{\rm pixon}} 
W(\delta(\bfx),\delta_l)[((F \otimes B)\times V_l) \otimes P_l](\bfx).
\label{appeq:easy}
\end{equation}
$V_l$ represents a mask that is unity in pixels with $\delta=\delta_l$
and zero otherwise. The calculation can be seen to be a series of
$n_{\rm pixon}$ correlations, hence the $n \log n$ scaling. $F$ is
readily shown to be $-2R/(n\sigma^2)$ and
$-2(D+\min(D,1)-\that*B)/(n(D+1))$ for $\chi^2$ and $\chi_\gamma^2$
respectively. 

In the case of the autocorrelation of the residuals
when the anticipated noise amplitude is independent of signal and
position in the image (\ie for
the examples considered in Section~\ref{ssec:sz}), equation
(\ref{appeq:easy}) is still applicable, but the derivative of $E_R$
with respect to $(\that*B)(\bfx)$ needs to include a sum over lag terms:
\begin{equation}
F(\bfx)=\frac{-2}{n\sigma^2}\sum_\bfz A_R(\bfz) (R(\bfx+\bfz)+R(\bfx-\bfz)).
\label{appeq:easyer}
\end{equation}
The sum of residuals comes from the derivative of $A_R(\bfz)$ with
respect to $R(\bfx)$.

When the anticipated noise amplitude $\sigma$ also depends upon the
signal in the pixel, as is the case for the Poisson noise used in the
examples in Section~\ref{ssec:xray}, the derivative calculation
becomes somewhat more involved. Referring back to equation
(\ref{appeq:easy}), the required form for $F$ becomes
\begin{equation}
F(\bfx)=\frac{-2}{n}\sum_\bfz A_R(\bfz) K(\bfx)(L(\bfx+\bfz)+L(\bfx-\bfz)).
\label{appeq:harder}
\end{equation}
where
\begin{equation}
K(\bfx)=\frac{D(\bfx)-0.5R(\bfx)}{(\that*B)(\bfx)^{1.5}}
\label{appeq:kdef}
\end{equation}
and
\begin{equation}
L(\bfx)=\frac{R(\bfx)}{\sqrt{(\that*B)(\bfx)}}.
\label{appeq:ldef}
\end{equation}
\end{appendix}


\begin{thebibliography}{}

\bibitem{betal} Bunn E., Fisher K.B., Hoffman Y., Lahav O., Silk J.,
Zaroubi S., 1994, ApJ, 432, L75

\bibitem{cff} Cavaliere A., Fusco-Femiano R., 1976, A\&A, 49, 137

\bibitem{detal1} Dixon D.D. et al., 1996, A\&AS, 120, 683

\bibitem{detal2} Dixon D.D. et al., 1997, ApJ, 484, 891

\bibitem{enf} Eke V.R., Navarro J.F., Frenk C.S., 1998, ApJ, 503, 569

\bibitem{fetal} Fisher K.B., Lahav O., Hoffman Y., Lynden-Bell D.,
Zaroubi S., 1995, MNRAS, 272, 885

\bibitem{g} Gull S.F., 1989, in Skilling J., ed., Maximum Entropy and
Bayesian Methods.  Kluwer, Dordrecht, p. 53

\bibitem{hjlb} Hobson M.P., Jones A.W., Lasenby A.N., Bouchet F.R.,
1998, MNRAS, 300, 1

\bibitem{ketal} Kn\"odlseder J. et al., 1996, SPIE Vol. 2806, 386

\bibitem{letal} Lahav O., Fisher K.B., Hoffman Y., Scharf C., Zaroubi
S., 1994, ApJ, 423, L93

\bibitem{mhkpp} Metcalf T.R., Hudson H.S., Kosugi T., Puetter R.C.,
Pi\~na R.K., 1996, ApJ, 466, 585

\bibitem{m} Mighell K.J., 1999, ApJ, 518, 380

\bibitem{pp1} Pi\~na R.K., Puetter R.C., 1992, PASP, 104, 1096 (PP92)

\bibitem{pp2} Pi\~na R.K., Puetter R.C., 1993, PASP, 105, 630 (PP93)

\bibitem{p} Puetter R.C., 1996, SPIE Vol. 2827

\bibitem{py} Puetter R.C., Yahil A., 1999, to appear in Mehringer D.,
Plante R., \& Roberts D., eds. ASP Conf. Ser., Astronomical Data
Analysis Software and Systems VIII. San Francisco (astro-ph/9901063)

\bibitem{ps}
Press W.H., Schechter P., 1974, ApJ, 187, 425

\bibitem{numrec}
Press W.H., Teukolsky S.A., Vetterling W.T., Flannery B.P., 1992,
Numerical Recipes, Cambridge University Press

\bibitem{rp} Rybicki G.B., Press W.H., 1992, ApJ, 398, 169

\bibitem{cmbfast} Seljak U., Zaldarriaga M., 1996, ApJ, 469, 437

\bibitem{s}
Skilling J., 1989, in Skilling J., ed., Maximum Entropy and Bayesian Methods.
Kluwer, Dordrecht, p. 45

\bibitem{wavelets}
Slezak E., Bijaoui A., Mars G., 1990, A\&A, 227, 301

\bibitem{samrpp}
Smith C.H., Aitken D.K., Moore T.J.T., Roche P.F., Puetter R.C.,
Pi\~na R.K., 1995, MNRAS, 273, 354

\bibitem{sz} Sunyaev R.A., Zel'dovich Ya. B., 1972,
Comm. Astrophys. Space Phys., 4, 173

\bibitem{cobrassamba}
Tauber J., Pace O., Volont\'e S., 1994, ESAJ, 18, 239

\bibitem{w}
Weiner N., 1949, in Extrapolation and Smoothing of Stationary Time
Series. Wiley, New York

\end{thebibliography}
\end{document}